\documentclass{iucrjournals}

\usepackage{multirow} 
\usepackage{amsfonts}
\usepackage{rotating} 

\title{Optimizing MHz X-ray photon correlation spectroscopy with burst-mode free-electron laser operation}

\author[a]{Johannes Möller\IUCrCemaillink{johannes.moeller@xfel.eu}\IUCrOrcidlink{0000-0001-8363-9077}}%

\author[a]{Wonhyuk Jo\IUCrAufn{Current affiliation: Korean Basic Science Institute, 28119 Cheongju-si, Korea }}%
\author[a]{Trey Guest\IUCrOrcidlink{0000-0001-7222-0588}}
\author[c]{Lars Fröhlich}
\author[c]{Matthias Scholz}
\author[c]{Marc Guetg}
\author[a]{Felix Brauße}
\author[a, b]{Aliaksandr Leonau}
\author[a]{James Wrigley}
\author[a]{Ulrike Boesenberg}
\author[a]{Lea Brings}
\author[a]{Jörg Hallmann\IUCrOrcidlink{0000-0002-1996-397X}}
\author[a]{Jan-Etienne Pudell\IUCrOrcidlink{0000-0001-8268-444X}}
\author[a]{Angel Rodriguez-Fernandez\IUCrAufn{Current affiliation: Diamond Light Source, Didcot, Oxfordshire OX11 0DE, United Kingdom }\IUCrOrcidlink{0000-0003-0587-5263}}
\author[a]{Roman Shayduk\IUCrOrcidlink{0000-0002-3577-8918}}
\author[a]{Alexey Zozulya\IUCrOrcidlink{0000-0003-1048-3432}}

\author[b,c]{Nimmi Das Anthuparambil\IUCrAufn{Current affiliation: Department of Physics, IIT Madras, Chennai, Tamil Nadu 600036, India}\IUCrOrcidlink{0000-0001-5338-2004}}
\author[b]{Michelle Dargasz\IUCrOrcidlink{0009-0008-1925-9860}}
\author[b]{Sonja Timmermann\IUCrOrcidlink{0000-0002-5428-9044}}
\author[d]{Maddalena Bin\IUCrOrcidlink{0000-0003-4906-9335}}
\author[d]{Foivos Perakis\IUCrOrcidlink{0000-0001-9863-9811}}
\author[e]{Michael Paulus\IUCrOrcidlink{0000-0002-3409-6798}}
\author[c]{Felix Lehmkühler\IUCrOrcidlink{0000-0003-1289-995X}}
\author[b]{Christian Gutt\IUCrOrcidlink{0000-0002-0051-8542}}
\author[a]{Anders Madsen\IUCrOrcidlink{0000-0001-6594-1029}}

\affil[a]{European X-ray Free-Electron Laser Facility, 22869 Schenefeld, Germany}
\affil[b]{Department Physik, Universität Siegen, 57072 Siegen, Germany}
\affil[c]{Deutsches Elektronen-Synchrotron DESY, 22603 Hamburg, Germany}
\affil[d]{Department of Physics, AlbaNova University Center, Stockholm University, 10691 Stockholm, Sweden}
\affil[e]{Fakultät Physik/ Dortmunder Elektronenspeicherringanlage, Technical University of Dortmund, 44221 Dortmund, Germany}

\begin{document} 
\nolinenumbers
\maketitle 

\begin{synopsis}
We develop a signal-to-noise model for MHz X-ray photon correlation spectroscopy (XPCS) at burst-mode XFELs in the dose-limited regime and show why fewer, more intense pulses can improve the signal-to-noise ratio. An experimentally demonstrated rolling pulse pattern enables substantially higher per-pulse intensities while preserving equilibrium dynamics, providing a practical strategy for dose-efficient XPCS.
\end{synopsis}

\begin{abstract}
Burst-mode X-ray free-electron lasers (XFELs) provide a powerful platform for X-ray photon correlation spectroscopy (XPCS) experiments with sub-microsecond temporal resolution, but the achievable performance is constrained in practice by both photon statistics and the sample's dose tolerance. Here, we develop an analytical framework for estimating the signal-to-noise ratio (SNR) of pulsed XPCS measurements, including pulse-to-pulse fluctuations of the incident intensity. We combine this description with dose constraints and show that, for a fixed maximum dose, acquiring fewer frames at higher per-pulse intensity yields a higher SNR than distributing the dose over a larger number of low-intensity frames. The framework is validated against measurements at the European XFEL, where the intensity fluctuations of the investigated operation mode are found to cause only a negligible reduction in SNR. We further introduce and experimentally demonstrate a rolling bunch pattern with three pulses per train and variable pulse spacings. This scheme enables more than an order-of-magnitude increase in the accessible per-pulse intensity while maintaining equilibrium dynamics. Our results provide practical guidelines for optimizing dose-efficient MHz-XPCS experiments and for evaluating future XFEL operation modes.
\end{abstract}


\section{Introduction}

X-ray photon correlation spectroscopy (XPCS) is a coherent-scattering technique for measuring structural dynamics over nanometer to atomic length scales \cite{madsen2020, lehmkuehler2021}. By correlating scattered speckle intensities at different times, XPCS provides access to the intermediate scattering function and hence to relaxation processes such as diffusion, caging, aggregation, and structural rearrangement. The short X-ray wavelength further enables measurements at scattering vectors that are inaccessible to conventional dynamic light scattering, making XPCS particularly useful for concentrated or opaque soft-matter and biological systems.

The availability of high-repetition-rate X-ray free-electron lasers (XFELs) has extended XPCS applications toward time scales that are difficult to impossible to access at storage-ring sources. Early XFEL-XPCS experiments demonstrated the feasibility of correlating speckle patterns from individual pulses at low repetition rates \cite{carnis2014,lehmkuehler2015,lehmkuhler2018}. The development of superconducting-linac-based facilities, in particular the European XFEL, subsequently enabled sequential XPCS at megahertz (MHz) intra-train repetition rates \cite{lehmkuhler2020,madsen2021}. In this regime, successive femtosecond X-ray pulses provide a sequence of structural snapshots with sub-microsecond temporal separation, while the pulse trains are repeated at a much lower rate that allows for sample translation \cite{dallari2021analysis, leonau2025}.

Taking advantage of burst-mode operation, MHz-XPCS has since been applied to a range of soft-matter and biologically relevant systems. Experiments have revealed beam-heating-modulated diffusion and non-equilibrium dynamics in nanoparticle suspensions \cite{lehmkuhler2020,dallari2021microsecond}, while related pump-probe strategies have used the X-ray beam itself to trigger and follow rapid structural transformations \cite{dallari2024}. Measurements on antibody and protein solutions have further shown that equilibrium dynamics can be accessed before radiation-induced aggregation becomes dominant, motivating the concept of ``correlation before aggregation'' \cite{reiser2022}. More recent studies of crowded protein solutions and complex colloidal systems have used MHz-XPCS to investigate caging, collective diffusion, self-assembly, and crystallization dynamics \cite{girelli2025,dargasz2025, anthuparambil2025, jo2025}.

Despite these advances, XPCS at a pulsed XFEL is not simply a faster version of XPCS at a quasi-continuous source. At a synchrotron radiation source, the measurement statistics are commonly described in terms of a continuous photon flux and an integration time, whereas at an XFEL the data are acquired as a sequence of individual pulses. The number of pulses per train as well as the number of trains, the photons detected per pulse and the pulse-to-pulse intensity fluctuations all influence the achievable signal-to-noise ratio. At the same time, the high peak intensity of XFEL pulses can make radiation damage the primary experimental constraint, particularly for soft and biological materials. The discrete source structure therefore provides not only access to shorter delays, but also opportunities for flexible allocation of dose across space and time.

For many MHz-XPCS experiments, the limiting factor is also not the available photon flux but the radiation tolerance of the investigated sample. Soft-matter systems are particularly susceptible to beam-induced effects, aggregation, or structural changes, which can alter the intrinsic dynamics and restrict the usable exposure time and pulse intensity. The relevant optimization problem is consequently not simply to maximize the number of detected photons, but to maximize the information obtained per absorbed dose \cite{moeller2019b, moller2021}.

In this work, we derive an analytical expression for the signal-to-noise ratio of XPCS at pulsed XFEL sources, explicitly accounting for the discrete pulse structure and pulse-to-pulse intensity fluctuations. We combine this description with an absorbed-dose constraint to establish scaling relations for the optimization of pulse intensity and the number of exposures. The resulting predictions are compared with experimental data and used to evaluate dose-limited acquisition strategies at the European XFEL. In particular, we investigate how the flexible pulse structure of burst-mode operation can be used to redistribute the dose across time points and compare conventional sequential acquisition with a variable, \textit{rolling} pulse pattern. Our results demonstrate the potential of pulse-pattern optimization to improve the per-pair signal-to-noise ratio and reduce data volumes, while parameterizing these improvements with respect to scattering strength, sample tolerance, temporal coverage, and repetition rate.

\section{Signal-to-Noise Ratio and Dose in MHz-XPCS}
\label{sec:snr}

\subsection{X-ray Photon Correlation Spectroscopy}

XPCS measures sample dynamics through correlations of the time-dependent speckle intensity. Let $I(\vec{q}, t)$ be the scattered intensity at scattering vector $\vec{q}$ and time $t$. The normalized two-time correlation function (TTCF) expresses the correlation between the intensity at two times as

\begin{equation}
TTCF(q, t_1, t_2) = \frac{\langle  I(\vec{q},t_1) I(\vec{q},t_2) \rangle_q}{\langle  I(\vec{q},t_1) \rangle_q \langle I(\vec{q},t_2) \rangle_q},
\end{equation}
with $\langle  ... \rangle_q$ denoting the ensemble average over detector pixels corresponding to a given range of scattering vectors. For the case of stationary systems, the correlation depends only on the time difference between measurements
\begin{equation}
\tau = t_2 -t_1,
\end{equation}
and the intensity auto-correlation can be written as 
\begin{equation}
    g_2(q,\tau) = \frac{\langle I(q,t) I(q,t+\tau)\rangle}{\langle I(q,t)\rangle^2},
\end{equation}
where $\langle  ... \rangle$ denotes the averaging over detector pixels $q$ and times $t$.
This general formulation applies to both quasi-continuous and pulsed X-ray sources. The intensity auto-correlation function provides access to microscopic dynamics through the Siegert relation
\begin{equation}
    g_2(q, \tau) = 1+ \beta(q) | f(q,\tau) |^2,
\end{equation}
where $f(q,\tau)$ is the intermediate scattering function. It is defined as $f(q, \tau) = S(q, \tau)/S(q, 0)$, with $S(q,\tau)$ being the time-dependent structure factor. The normalized ISF $f(q,\tau)$ quantifies how the sample structure changes over a delay $\tau$ at length scales of approximately $2\pi/q$.

The speckle contrast $\beta(q)$ sets the amplitude of the intensity correlation above its baseline, and is therefore an important quantity determining the SNR of a measurement. It approaches unity for fully coherent illumination and fully resolved speckles, and is reduced by partial coherence and finite detector resolution \cite{moeller2019b}. The decay of $| f(q,\tau) |^2$ yields characteristic relaxation times, while the $q$ dependence provides information about the underlying motion. For example, simple Brownian diffusion gives $f(q, \tau) = \exp( - D q^2 \tau)$, allowing the diffusion coefficient $D$ to be determined.

\subsection{Signal-to-noise ratio}
In the photon-counting limit, the variance of the correlation function $g_2$ can be written as \cite{jakeman1971, falus2006}
\begin{equation}
    \mathrm{Var} (g_2) = g_2/(N \cdot \langle k \rangle^2),
\end{equation}
with $N$ being the total number of correlated intensity pairs averaged to yield $g_2$ and $\langle k \rangle$ the mean number of detected photons per acquisition and detector pixel. For a quasi-continuous source such as a synchrotron radiation source, and neglecting detector dead time and waiting times between successive frames, the integration time per acquisition is equal to the lag time $\tau$ between neighboring frames. The mean number of detected photons per pixel and acquisition is therefore given by the product of the mean count rate per pixel $I$ and $\tau$:
\begin{equation}
    \langle k \rangle = I \cdot \tau.
    \label{eq:k_cont}
\end{equation}

The total number of correlated intensity pairs is determined by the number of frames $N_{fr} = T / \tau$, with $T$ being the total measurement time, and $N_{pix}$ the number of pixel within the considered region of interest on the detector:
\begin{equation}
    N = N_{fr} \cdot N_{pix} = T / \tau \cdot N_{pix} \label{eq:N_cont}
\end{equation}
For a continuous source, this leads to the expression
\begin{equation}
    \mathrm{Var}(g_2) = g_2/(T \cdot \tau \cdot N_{pix} \cdot  I ^2), \label{eq:synchrotron}
\end{equation}
from which the well known estimation for the SNR follows as
\begin{equation}
    SNR = \frac{g_2 -1}{\sqrt{{\mathrm{Var} \ g_2}}} = \frac{g_2 -1}{\sqrt{g_2}} \cdot I  \cdot\sqrt{(T \cdot \tau \cdot N_{pix})} \approx \beta \cdot  I   \cdot\sqrt{(T \cdot \tau \cdot N_{pix})} .
\end{equation}

This formulation assumes that we can estimate $g_2$ with its zero time limit $1+\beta$ and that the contrast $\beta$ is rather small, which is true for typical XPCS measurements.
In order to transfer this calculation to XFELs, some underlying assumptions must be revisited.

\subsection{Pulsed X-ray sources}

In the quasi-continuous case, the total measurement time $T$ is typically much longer than the acquisition time $\tau$, such that many frames contribute to the correlation function. The number of frame pairs is then $N \propto N_{fr} = T/\tau$, where $N_{fr} \gg 100$. This scaling no longer holds for pulsed sources, where measurements often comprise far fewer pulses, down to double shot experiments with $N_{fr} = 2$. In this limit, only a single frame pair contributes and eq. \ref{eq:N_cont} must be modified accordingly to

\begin{equation}
    N = (N_{fr}-1)  \cdot N_{pix} \cdot N_{rep}. \label{eq:N_puls}
\end{equation}

Additionally, the temporal sampling is set by the repetition rate of the source, and not necessarily anymore by an integration time. This means that $\tau$ now describes a delay or lag time between pulses and not an integration time anymore. Consequently, the mean number of detected photons per acquisition $\langle k \rangle_i$ is determined by the incident intensity of the individual pulse and is independent of $\tau$.

At the European XFEL, XPCS measurements are performed in burst mode, where X-ray pulses are delivered in trains separated by relatively long gaps \cite{decking2020mhz}. Within each train, pulses arrive at MHz repetition rates, while the trains repeat at much lower frequency of $10$ Hz. In the context of XPCS analysis, each pulse train produces an independent intensity time series, from which a single correlation function $g_2 (q, \tau)$ can be calculated. In this framework, $N_{fr}$ corresponds to the number of pulses within a single train, while $N_{rep}$ denotes the number of pulse trains.

\subsection{Fluctuations of the Incident Intensity}
So far, the established consideration also assumes that each frame detects nearly the same average number of photons. This assumption is certainly a good approximation for synchrotron radiation sources. However, at XFELs, shot-to-shot intensity fluctuations are an intrinsic phenomenon. Cao et al.~\cite{cao2020} calculated numerically that these pulse-to-pulse intensity fluctuations can reduce the experimental SNR significantly. We hereafter include this effect in the analytical estimation of the SNR.

We consider a series of measured intensity distributions $I_i$, where $I_i$ denotes the intensities recorded in frame $i$. For neighboring frames separated by a delay $\tau_1$, the averaged correlation is obtained as 

\begin{equation}
   \bar{g_2}(\tau_1) = {1 \over N_{fr} -1} \sum_i^{N_{fr}-1} g_2^i(\tau_1)= {1 \over N_{fr} -1} \sum_i^{N_{fr}-1} \frac{ \langle I_i I_{i+1} \rangle}{\langle I_i \rangle \langle I_{i+1} \rangle},\label{eq:g2_pulsed}
\end{equation}
where $\langle ... \rangle$ denotes the averaging over the individual pixel values of each frame. Therefore, the mean intensity of all frames is denoted
\begin{equation}
    \langle \bar{k} \rangle = {1 \over N_{fr}} \sum_i^{N_{fr}} \langle k \rangle_i ,
\end{equation}
with $\langle k \rangle_i = \langle I_i \rangle$.
In the low intensity regime, sparse data ($\langle \bar{k} \rangle < 1$), the variance of $\bar{g_2}$ is dominated by Poisson statistics of the individual frames, so that \cite{jakeman1971, falus2006}
\begin{equation}
    \mathrm{Var}(g_2) = {\bar{g_2} \over N_{pix} \langle \bar{k}\rangle^2}.\label{eq:var}
\end{equation}
This expression is equivalent to eq. \ref{eq:synchrotron}, which assumes, however, that the mean intensity in each frame is the same $ \langle k \rangle_i \approx \langle \bar{k} \rangle$, i.e. $\mathrm{Var}(\langle k \rangle_i) \to 0$.

This is not a valid assumption for XFEL-based XPCS and we therefore expand eq. \ref{eq:var} by explicitly taking into account possible fluctuations of the incident beam intensity, which result in fluctuations of the mean scattering intensity $\langle k \rangle_i$. We decompose the variance of the correlation function to
\begin{equation}
    \mathrm{Var}(g_2^i) = \mathbb{E}_{\langle k \rangle}[\mathrm{Var}(g_2^i | \langle k \rangle_i , \langle k \rangle_{i+1})] + \mathrm{Var}_{\langle k \rangle}(\mathbb{E}[g_2^i | \langle k \rangle_i , \langle k \rangle_{i+1}]).
\end{equation}
Since the correlation function defined in eq. \ref{eq:g2_pulsed} is normalized, it is independent of the actual value of $\langle k \rangle_i$ and $\langle k \rangle_{i+1}$, so that
\begin{equation}
    \mathbb{E}[g_2^i | \langle k \rangle_i, \langle k \rangle_{i+1}]  = \bar{g_2},
\end{equation}
and therefore
\begin{equation}
    \mathrm{Var}(\mathbb{E}[g_2^i | \langle k \rangle_i, \langle k \rangle_{i+1}])  = 0.
\end{equation}
This leads to
\begin{equation}
    \mathrm{Var}(g_2^i) = \mathbb{E}_{\langle k \rangle}[\mathrm{Var}(g_2^i | \langle k \rangle_i ,\langle k \rangle_{i+1})] = \mathbb{E}_{\langle k \rangle}\left[ \frac{\bar{g_2}}{N_{pix} \langle k \rangle_i \langle k \rangle_{i+1}} \right] = \frac{\bar{g_2}}{N_{pix} } \mathbb{E}_{\langle k \rangle} \left[ \frac{1}{\langle k \rangle_i \langle k \rangle_{i+1}} \right].
\end{equation}

Next, we assume that the distribution of $\langle k \rangle_i$ follows a gamma distribution \cite{geloni2010} with the number of modes $M$ and the scale $\Theta = \langle \bar{k} \rangle/M$, and that $\langle k \rangle_i$ and $\langle k \rangle_{i+1}$ are independent samples of this distribution. It follows that
\begin{equation}
    \mathbb{E} \left[ \frac{1}{\langle k \rangle_i \langle k \rangle_{i+1}} \right] =  \left( \mathbb{E} \left[ \frac{1}{\langle k \rangle_i} \right] \right)^2 =  \left( \frac{M}{\langle \bar{k} \rangle (M-1)} \right)^2 =  \frac{1}{\langle \bar{k} \rangle^2} \frac{1}{(1 - CV^2)^2} ,
\end{equation}
for $M >1$ and with $CV = 1/\sqrt{M}$ being the normalized standard variation or coefficient of variation
\begin{equation}
    CV =   \frac{1}{\bar{ \langle k \rangle}} \sqrt{\frac{\sum_{i=1}^{N_{fr}} ( \langle k \rangle_i- \langle \bar{k} \rangle)^2   }{  {N_{fr}}}} .
\end{equation}

This leads to the expression

\begin{equation}
    \mathrm{Var}(\bar{g}_2) = \frac{\bar{g_2}}{N_{pix} \cdot \bar{ \langle k \rangle}^2 \cdot (1- \mathrm{CV}^2)^2}
\end{equation}

and the overall SNR can be estimated as

\begin{equation}
     SNR = \frac{\bar{g_2} -1}{\sqrt{{\mathrm{Var} (\bar{g}_2)}}} \approx \beta \cdot \bar{\langle k \rangle} \cdot (1-CV^2) \cdot\sqrt{(N_{fr}-1) \cdot N_{rep} \cdot N_{pix}}; \ \mathrm{for} \ CV<1. \label{eq:snr_pulsed}
\end{equation}
This equation describes the SNR of multispeckle XPCS at pulsed, fluctuating sources. A numerical validation is shown in Appendix \ref{sec:model}. Importantly, the variance expression becomes singular at $CV = 1$, while the corresponding SNR tends to zero as $CV \to 1$. This can be easily understood by the fact that broad distributions ($CV \geq 1$) include a finite probability to record frames with $\langle k \rangle_i = 0$, which makes eq. \ref{eq:g2_pulsed} diverge, and that low intensity frames dominate the SNR. In practice, this can be avoided by truncating the measured ensemble to $\langle k \rangle_i > \langle k \rangle_{min}$ and introducing a weighted average in eq. \ref{eq:g2_pulsed} \cite{cao2020}. This, however, may also introduce a bias towards beam driven dynamics, with high intensity frames being more susceptible to this effect.

\subsection{Temporal sampling and experimental verification}
\label{sec:temp}

So far, the explicit time dependence of the intensity autocorrelation function $g_2(\tau)$ has been ignored in the SNR considerations. We now include this by noting that $g_2$ is evaluated at delay times $\tau_{n} = n/f$, where $f$ denotes the source and detector repetition rate and $n$ is the frame-index separation between two detector frames measured at $t_i$ and $t_{i+n}$, with $1 \le n \le N_{fr}-1$.

\begin{figure}[h]
\begin{center}
\includegraphics[width=0.75\textwidth]{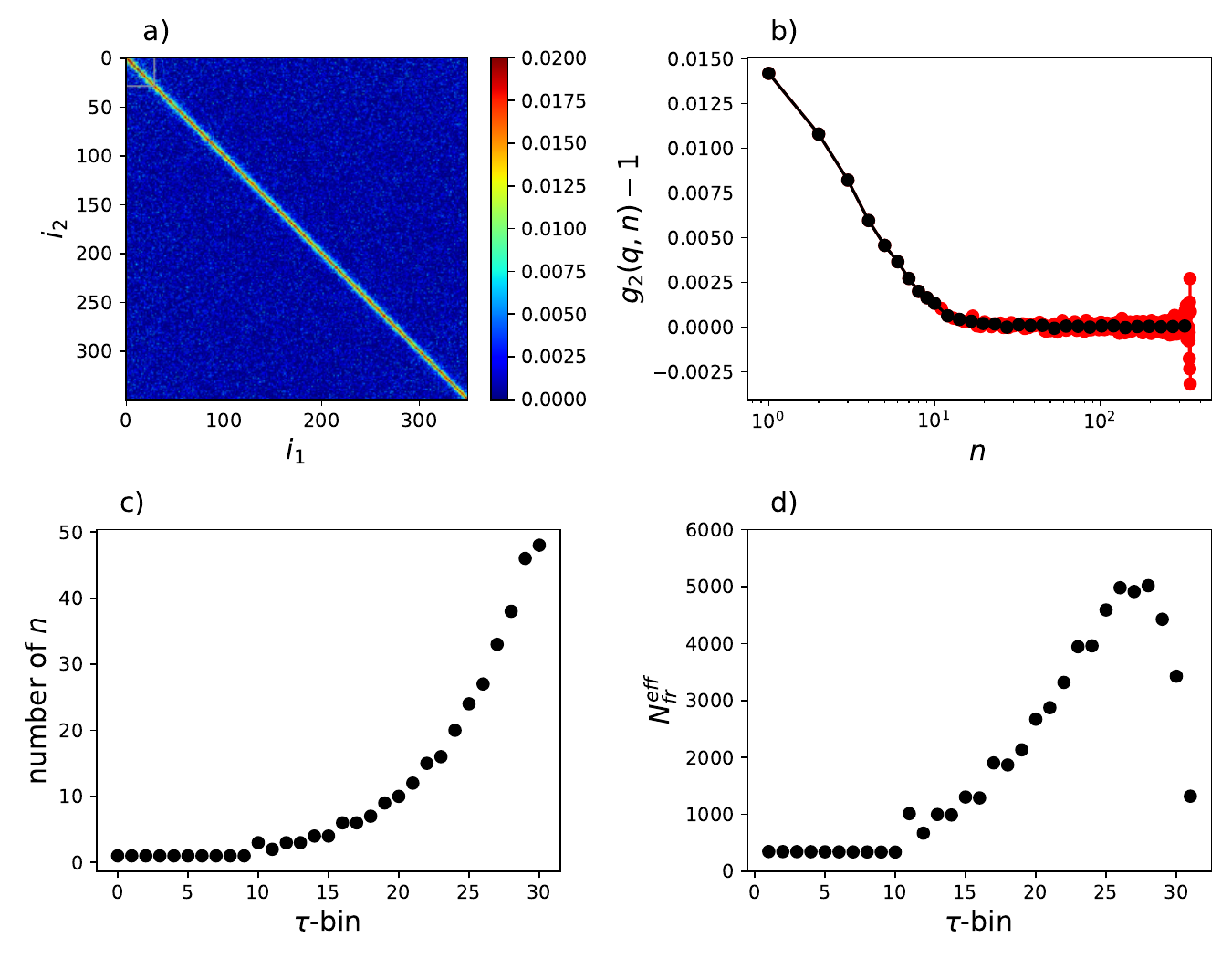}
\end{center}
\caption{a) Two-time correlation function (TTCF) of silica nanoparticles in water, measured at the MID instrument of the European XFEL. b) Intensity autocorrelation function $g_2(q,n)$ extracted from a). The red curve shows the directly computed correlation, while the black curve represents the same data after temporal log-binning of the delay-time axis. c) Number of TTCF time points (corresponding to the red data points in b)) contributing to each time bin (black points in b)). d) Effective number of frame pairs, $N_{fr}^{eff}$, per time bin. The increase in $N_{fr}^{eff}$ at longer delay times leads to an improved signal-to-noise ratio for these data points. \label{fig:silica}}
\end{figure}

An example measurement acquired at the MID instrument of the European XFEL on a silica-nanoparticle suspension is shown in Fig.~\ref{fig:silica}. Details of the experimental setup are listed in Appendix \ref{sec:exp}. Panel~(a) shows the TTCF evaluated at $q = 0.125~\mathrm{nm}^{-1}$ using $N_{\mathrm{fr}}=350$ frames per train, $N_{\mathrm{pix}}=2986$ detector pixels, and $N_{\mathrm{rep}}=6402$ repetitions (trains). The TTCF is reduced to the intensity autocorrelation function $g_2\!\left(q,n\right)$ by averaging all matrix elements with identical frame separation $n$; the resulting $g_2$ is shown as the red curve in panel~(b). A correction scheme specific to the detector is applied, in which the TTCF computed from neighboring trains is subtracted, as detailed in \cite{leonau2025}. This is required due to jumps of the darks pedestals and the quantity shown is therefore $TTCF_{corr} = TTCF_{\mathrm{on}}-TTCF_{\mathrm{off}}$, such that the baseline is near $0$ rather than $1$. In addition, pulse 30 is masked (white lines) due to detector artifacts in the corresponding storage cell of the AGIPD detector.

The previously used estimate of $N_{\mathrm{fr}}-1$ frame pairs is only valid for the shortest lag ($n=1$), which comprises entries directly next to the diagonal in Fig. \ref{fig:silica} a). For a general lag $n$, the number of available frame pairs decreases to $N_{\mathrm{fr}}-n$, i.e.\ each delay is sampled $N_{\mathrm{fr}}-n$ times. At longer delays, however, the correlation function is typically rebinned to a logarithmic time axis in order to improve statistics (black curve in Fig.~\ref{fig:silica}~(b)). As a result, many distinct lag times contribute to each logarithmic time bin. For example, when binning a correlation function measured with $N_{\mathrm{fr}}=350$ pulses into 30 logarithmically spaced bins, the final bin contains on the order of 50 different lag times (Fig.~\ref{fig:silica}~(c)). We therefore compute the effective number of frame pairs per time bin $N_{fr}^{eff}$, shown in Fig.~\ref{fig:silica}~(d). Despite the reduced sampling at long delays, temporal binning leads to a substantial improvement in SNR in this regime and reveals the non-uniform SNR across $\tau$.

\begin{figure}[h]

\begin{center}
\includegraphics[width=0.95\textwidth]{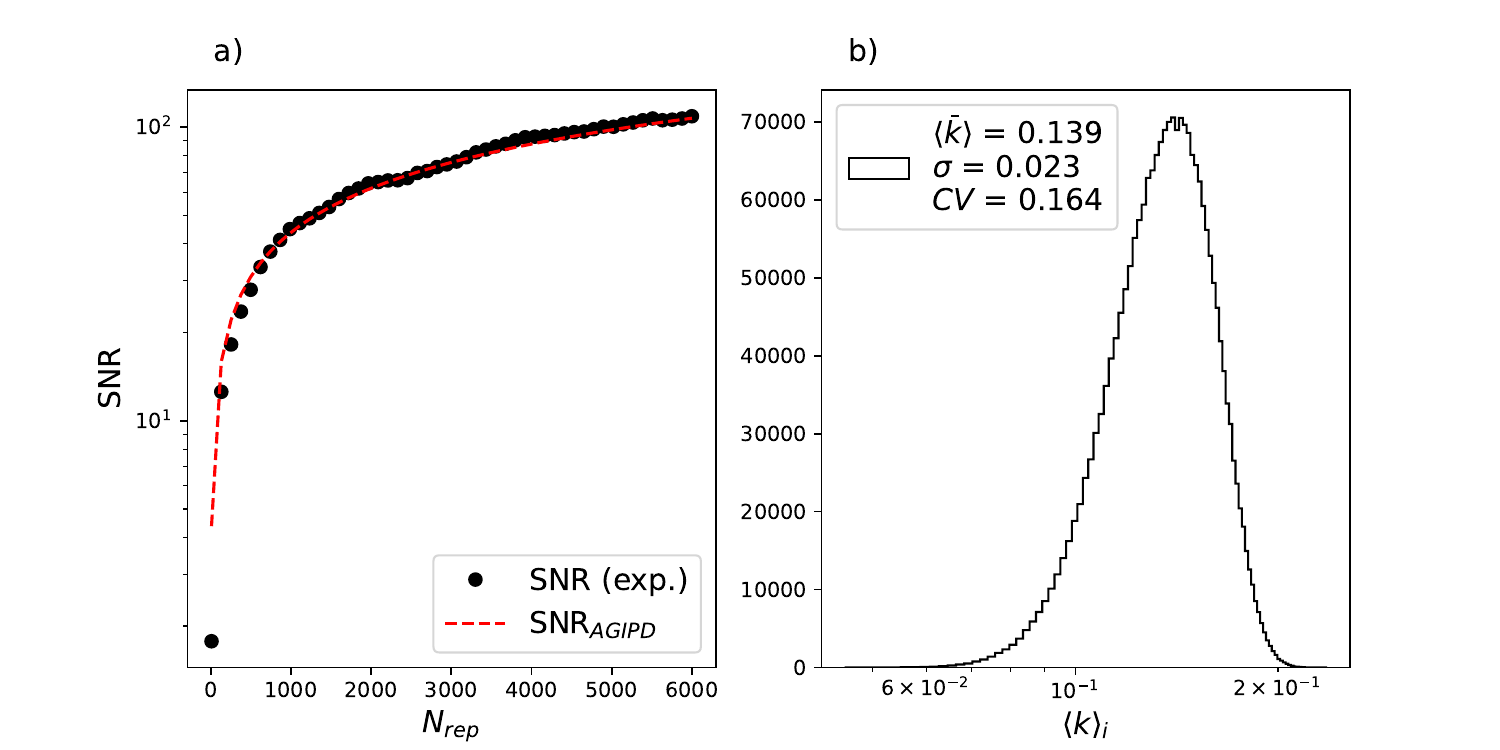}
\end{center}
\caption{a) Comparison between theoretical and experimentally achieved signal-to-noise ratio (SNR). Red line indicates the theoretical prediction based on Eqs. \ref{eq:snr_pulsed} and \ref{eq:snr_agipd}, while black circles represent the SNR calculated from the experimental data shown in Fig. \ref{fig:silica}.
b) Distribution of scattering intensities $\langle k \rangle_i$, shown as a histogram. The mean value, standard deviation and coefficient of variation of the distribution are indicated in the legend.\label{fig:snr}}
\end{figure}

Since the lowest SNR is obtained at the shortest lag time $\tau_{n=1}$, corresponding to neighboring frames, we compare the analytical and experimental results for this lag. We average the TTCF for different numbers of $N_{\mathrm{rep}}$ repetitions and extract only the first off-diagonal elements $g_2^{N_{rep}}(\tau_1) - 1 = \langle  TTCF_{corr} (i_2=i_1+1)\rangle_{N_{rep}}$. This yields $N_{\mathrm{fr}}-1=349$ statistically equivalent estimates of $g_2^{N_{rep}}(\tau_1)-1$, from which we compute
\begin{equation}
    \mathrm{SNR}_{\mathrm{exp}}(N_{\mathrm{rep}})
    = \frac{
    \langle \ g_2^{N_{rep}}(\tau_1) -1 \rangle_{fr}
    }
    {
    \mathrm{\sigma}_{fr}\!\left[g_2^{N_{rep}}(\tau_1)\right]
    } \sqrt{N_{fr}-1}.
    \label{eq:snr_exp}
\end{equation}
The subscript $fr$ indicates that the mean $\langle .. \rangle$ and standard deviation $\sigma$ are evaluated over the $349$ frame pairs. The resulting SNR is shown in Fig.~\ref{fig:snr} (a) as black dots. The corresponding analytical prediction (red line) is calculated from Eq.~\ref{eq:snr_pulsed} using $\beta \approx g_2-1 \approx 0.014$, $\bar{\langle k \rangle} = 0.139$ and $\mathrm{CV} = 0.164$. Due to the applied cross-correlation correction scheme for the AGIPD detector, the SNR calculation needs to be adapted by an additional factor:

\begin{equation}
    \mathrm{SNR}_{AGIPD} = \frac{g_2^{on} - g_2^{off}}{\sqrt{\mathrm{Var}(g_2^{on}- g_2^{off})}} \approx \frac{1}{\sqrt{2}} \cdot \mathrm{SNR}, \label{eq:snr_agipd}
\end{equation}
assuming $\mathrm{Var}(g_2^{on} - g_2^{off}) = \mathrm{Var}(g_2^{on}) + \mathrm{Var}(g_2^{off})$.

The analytical model expressed in eq.~\ref{eq:snr_pulsed} and \ref{eq:snr_agipd} reproduces the experimentally observed SNR. The intensity fluctuations are relatively small, as depicted in Fig.~\ref{fig:snr} (b), so that we only account a reduction factor of $(1- 0.16^2) = 0.97$ to this effect. Therefore, an SNR $>100$ can be reached in this case within 10 min of measurement time at a burst rate of 10 repetitions per second.

\subsection{Dose constraints}

In the previous section we derived an analytical description for the SNR and demonstrated that it describes the experimentally observed scaling, reaching close to the photon-statistics limit. Building on this description, we now extend the considerations by including experimental constraints that determine the accessible parameter space at FELs. In particular, we use the SNR expression to identify favorable operating conditions under limits set by a maximum tolerable radiation dose.

For a solution of scatterers (e.g.\ proteins in solution), the mean number of detected photons per pixel and frame at momentum transfer $q$ can be written as
\begin{equation}
    \langle \bar{k} \rangle (q) = \bar{\Phi}_0 \cdot T_{sample} \cdot d \cdot \frac{\mathrm{d} \Sigma}{\mathrm{d} \Omega}(q) \cdot \Delta \Omega_{pix},\label{eq:k_q}
\end{equation}
where $\bar{\Phi}_0$ is the average number of incident photons per pulse, $T_{\mathrm{sample}}$ is the sample transmission, and $d$ is the sample thickness. The quantity $\mathrm{d}\Sigma/\mathrm{d}\Omega(q)$ denotes the differential scattering cross section per unit volume (absolute scattering intensity). Finally, $\Delta\Omega_{\mathrm{pix}}=(P/L)^2$ is the solid angle subtended by a single detector pixel, with pixel size $P$ and sample-to-detector distance $L$.

The absorbed dose per sample position can be estimated as
\begin{equation}
    D = \frac{(N_{fr} - 1) \cdot \bar{\Phi}_0 \cdot (1-T_{sample}) \cdot E}{ d \cdot A \cdot \rho},\label{eq:dose}
\end{equation}
where $E$ denotes the photon energy, $\rho$ the density of the sample and $A$ the illuminated area on the sample.\footnote{The last pulse is not included in the dose sum, since its deposited energy is not probed by any subsequent frame.}
Assuming a maximum tolerable dose $D_{\max}$, Eq.~\ref{eq:dose} implies that the admissible pulse intensity (and thus $\bar{\Phi}_0$) decreases as $N_{\mathrm{fr}}$ increases. Combining Eq.~\ref{eq:k_q} with Eq.~\ref{eq:dose} therefore yields
\begin{equation}
    \langle \bar{k} \rangle (q) = K \cdot \frac{D_{max}}{N_{fr}-1} \cdot \frac{\mathrm{d} \Sigma}{\mathrm{d} \Omega}(q),\label{eq:k_dose_relation}
\end{equation}
with $K = \frac{d^2 \cdot A \cdot \rho \cdot \Delta \Omega_{pix}}{E} \cdot \frac{T_{sample}}{1 - T_{sample}}$.

\begin{figure}[ht]
\begin{center}
\includegraphics[width=0.95\textwidth]{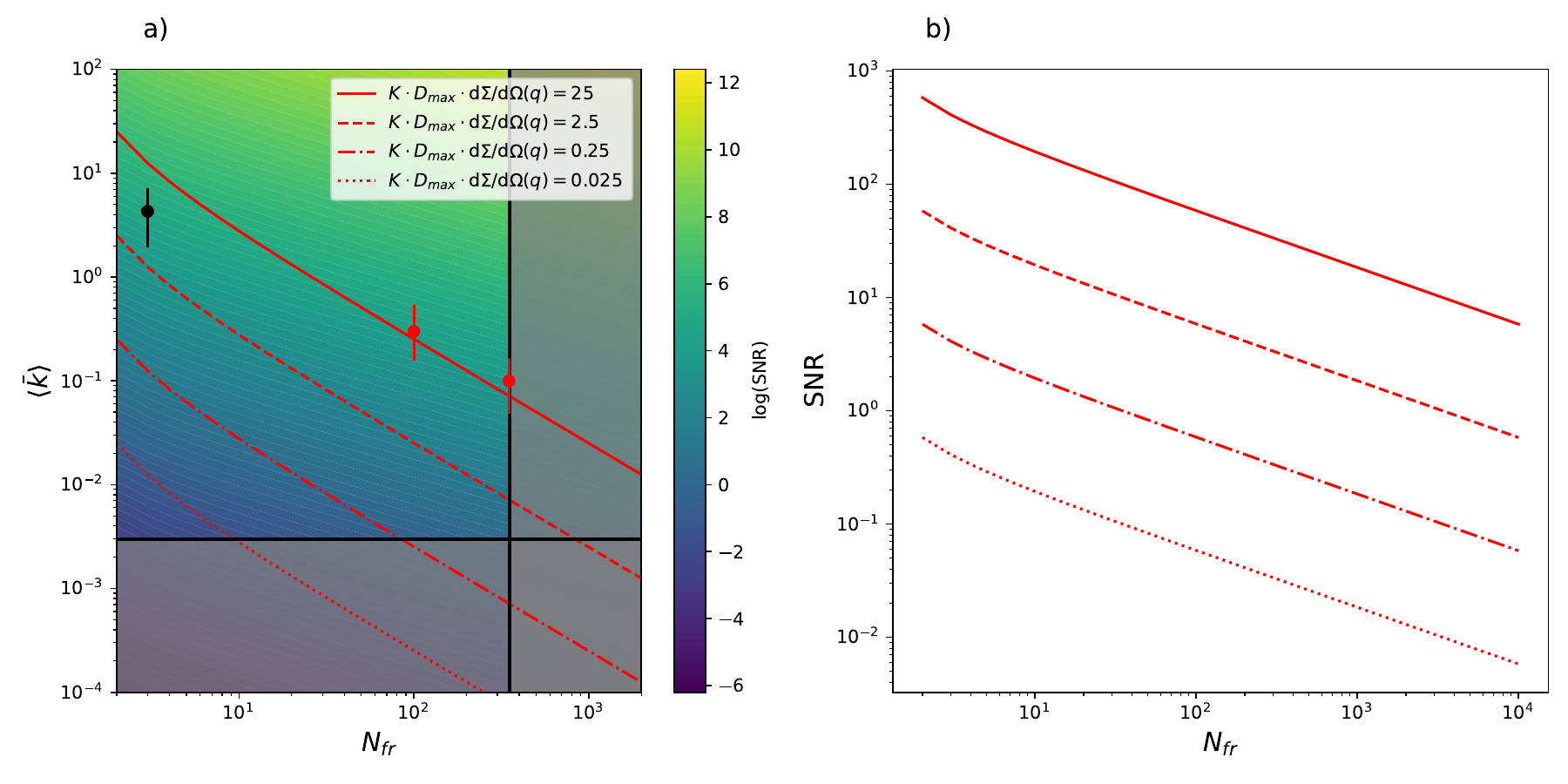}
\end{center}
\caption{a) Achievable SNR for different combinations on $\langle \bar{k} \rangle$ and $N_{fr}$. Red lines correspond to constant dose conditions, for different products of highest possible dose $D_{max}$ and scattering cross-section $\frac{\mathrm{d} \Sigma}{\mathrm{d} \Omega}(q)$. For measurements with the AGIPD, additionally a minimum average intensity of $\langle k \rangle < 0.003$ is required, and the maximum number of frames is limited to $N_{fr} = 351$. These constraints are indicated by black lines and the excluded regions are shaded. Black and red data points correspond to the experimental validation presented in sec. \ref{sec:pattern}. b) Same constant-dose lines as displayed in a), showing that the highest achievable SNR for each condition can be increased by decreasing the number of frames $N_{fr}$. \label{fig:snr_k_n} }
\end{figure}

Eq.~\ref{eq:k_dose_relation} is plotted in Fig.~\ref{fig:snr_k_n} a) for four values of the product $K\ \cdot D_{\max}\ \cdot \mathrm{d}\Sigma/\mathrm{d}\Omega(q)$ (red lines). Each red line represents the relationship between $\langle \bar{k}\rangle$ and $N_{fr}$ at constant dose. In practice, both values can be chosen freely within an experimentally accessible range\footnote{The upper bound in intensity is set by the unattenuated FEL pulse energy, while the maximum number of frames is limited by the available pulse pattern and/or the detector memory depth.}, but such that the chosen combination of $N_{\mathrm{fr}}$ and $\langle k\rangle$ remains below a certain red line (i.e., within a region of admissible dose depending on the sample characteristics).

The background color map shows the corresponding SNR for each combination of $\langle \bar{k}\rangle$ and $N_{fr}$, calculated from Eq.~\ref{eq:snr_pulsed} and \ref{eq:snr_agipd} with $\beta=1.4\%$, $N_{pix}=2000$, $N_{rep}=3000$, and $CV = 0.2$. While the parameters entering $K$ are largely fixed by the experimental setup, and can be regarded constant, $D_{max}$ and ${\mathrm{d} \Sigma}/{\mathrm{d} \Omega}(q)$ may vary substantially between samples. Consequently, higher radiation tolerance (large $D_{max}$) and/or stronger scattering (large ${\mathrm{d} \Sigma}/{\mathrm{d} \Omega}(q)$) translate directly into a higher achievable SNR.

Irrespective of the specific sample, Figs. \ref{fig:snr_k_n} a) and b) reveal that fewer frames acquired at higher per-pulse intensity outperform many low-intensity frames in terms of SNR. This result may seem counter-intuitive, as increasing the number of frames is commonly associated with improved noise reduction. In the dose-tolerance-limited regime, however, the opposite is true: fewer frames are better in terms of SNR.

\section{Optimizing MHz-XPCS using Variable Pulse Patterns}
\label{sec:pattern}

In the previous sections, we showed that A) the SNR of $g_2(\tau)$ varies strongly for different values of $\tau$ and B) reducing the total number of frames improves the SNR under dose-limited conditions. The question therefore arises how to choose the optimal number of frames for an experiment, taking into account potentially conflicting requirements to increase the SNR but also cover the required range of delay times $\tau$. We therefore set out to compare experimental data for different combinations of scattering intensity $\langle \bar{k} \rangle$ and number of frames $N_{fr}$. We additionally explore a rolling bunch pattern (see Appendix \ref{sec:roll} for implementation details), which offers the possibility to reduce the number of pulses per sample spot to $N_{fr} = 3$, without reducing the temporal range probed by $g_2(q, \tau)$, and potentially also provides a more uniform SNR distribution across delay times.

\subsection{Experimental demonstration of a rolling bunch pattern}\label{sec:rolling}

 In this scheme, the number of pulses per train is reduced to $N_{fr} = 3$, but the temporal spacing between them is varied from train to train. By cycling through ten distinct pulse delays, a wide range of $\tau$ can be probed, before repeating the sequence in order to build up statistics. This approach should allow for higher single-shot intensities and thus enhances the achievable SNR, as each sample position is only illuminated with three pulses. Additionally, detector related artifacts might be avoided or easier to mitigate, as detailed in Appendix \ref{sec:roll} as well.

\begin{figure}[ht]
\begin{center}
\includegraphics[width=0.95\textwidth]{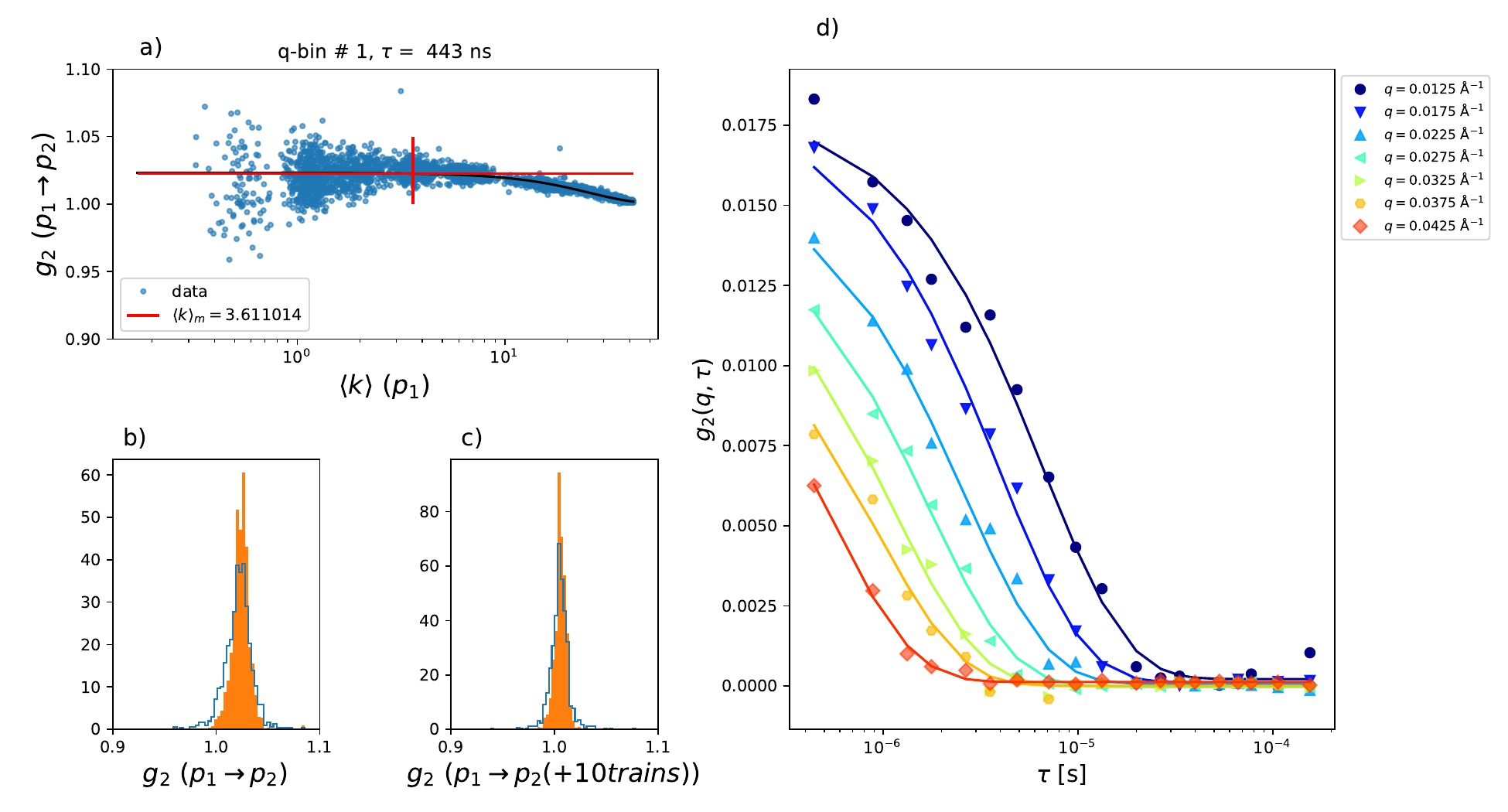}
\end{center}

\caption{Demonstration of the rolling bunch pattern on a sample of Ferritin in aqueous solution. a) Correlation between the first two pulses as a function of mean scattering intensity of the first pulse, shown only for the first $q$-bin and shortest lag time $\tau$. The onset of radiation damage can be seen by a decrease of correlation for higher intensities. The onset is marked as red vertical line and only values below the threshold are further considered. The determined threshold value is depicted as $\langle k \rangle_m$. b) Histogram (blue, intensity weighted orange) of the correlation values below damage threshold, as shown in a). c) Corresponding off-correlation values for baseline determination. d) Full correlation function constructed by merging all $q$-bins and lag times $\tau$. \label{fig:fer_rol}}
\end{figure}

Fig. \ref{fig:fer_rol} a) displays the single pulse-pair correlations $g_2(p_1, p_2)$ for pulses spaced by $\tau = 443 \ \mathrm{ns}$ within one specific $q$-bin, as a function of mean photon number of the first pulse $\langle k \rangle(p_1)$. The sample is the well studied case of ferritin in aqueous solution \cite{girelli2025}. $p_i$ marks the i-th pulse of each train. The incident intensity was changed by using solid attenuators with transmission factors ranging from $1.4 \cdot 10^{-4}$ to $1$ (no attenuation). As can be seen, the $g_2$ values are constant until a certain intensity threshold, above which the values start to decay. In order to obtain the dose threshold, the data points were fitted with an exponential function (black line)
\begin{equation}
    g_2 (\langle k \rangle) = g_2^0 \ \mathrm{exp}((- \langle k \rangle / \kappa)^\gamma)
    \label{eq:exp}
\end{equation}
and the threshold $\langle k \rangle_m$ is defined as the value at which $g_2 (\langle k \rangle)$ decayed to $0.975 \cdot g_2^0$. A histogram of all $g_2$ values with  $\langle k \rangle < \langle k \rangle_m = 3.61$ is shown in Fig. \ref{fig:fer_rol} b) in blue, and an intensity weighted histogram in orange. In addition, the off-correlation with the trailing train having the same pulse pattern (off by 10 trains) is provided in Fig. \ref{fig:fer_rol} c). From both distributions, the mean is calculated and the off-correlation value is subtracted to obtain one point of the correlation function. This procedure is repeated for all pulse-pairs and q-bins, which results in the correlation function displayed in Fig. \ref{fig:fer_rol} d). Notably, this correlation function is obtained with significantly higher single-frame intensities as studied using the conventional scheme \cite{girelli2025, dargasz2025}. However, the broad distribution of the intensities makes a quantified SNR discussion difficult.

\subsection{Scaling with dose limitations}

\begin{figure}[ht]

\begin{center}
\includegraphics[width=0.75\textwidth]{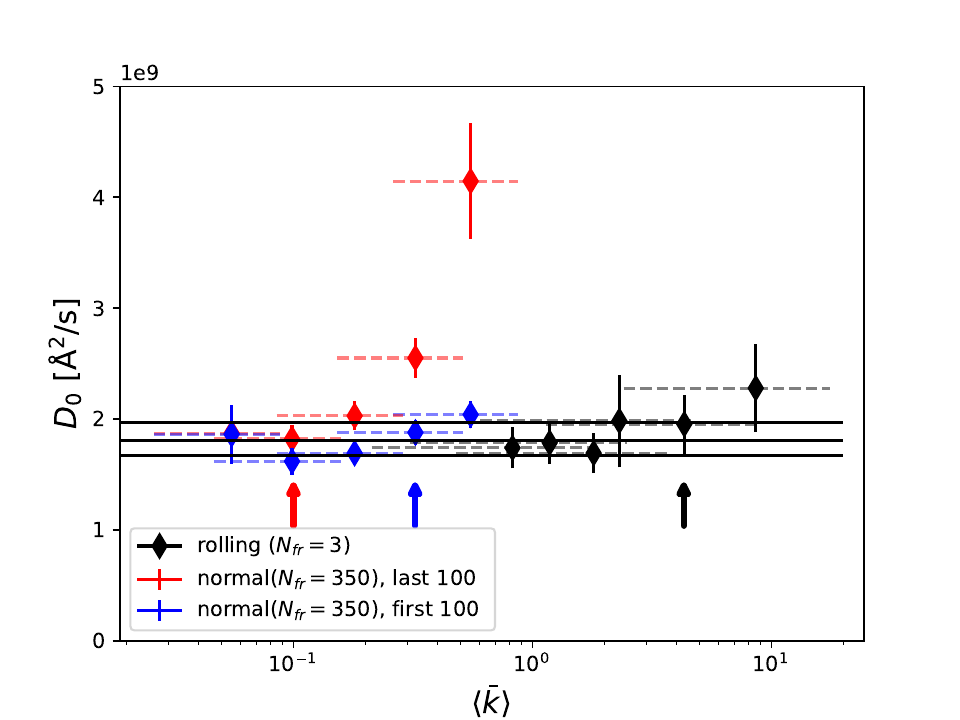}
\end{center}
\caption{Diffusion constant measured as a function of mean scattering intensity $\langle \bar{k} \rangle$ for the normal, sequential XPCS scheme with $N_{fr} = 350$ and the rolling bunch pattern scheme with $N_{fr} = 3$. Equilibrium diffusion constant is depicted as horizontal line as guide to the eye. The rolling bunch pattern allows accessing much higher scattering intensities while still maintaining equilibrium dynamics. The last data points within the ”equilibrium regime” for each condition are marked by arrows, which each correspond to the correlation functions shown in Fig. \ref{fig:snr_comp}.\label{fig:d0}}
\end{figure}

In order to assess possible improvements in terms of SNR, we turn to a sample which can be studied on a wide range of intensities, i.e. which corresponds to a large value of $K\ \cdot D_{\max}\ \cdot \mathrm{d}\Sigma/\mathrm{d}\Omega(q)$ as displayed in Fig. \ref{fig:snr_k_n}. We employ the same model system as in sec. \ref{sec:snr}, namely silica nanoparticles in water. Previous studies showed an increasing diffusion constant with increasing fluence, due to an X-ray induced heating of the sample \cite{lehmkuhler2020, dallari2021microsecond}. For different attenuator settings (i.e., for different values of $\bar{\langle k\rangle}$), we therefore determine the average diffusion constant $D_0$ by fitting the $q$-dependent relaxation rates extracted from the correlation functions with $\Gamma(q) = D_0 q^2$. The resulting $D_0$ as a function of $\bar{\langle k\rangle}$ is shown in Fig. \ref{fig:d0}, for a normal, sequential scheme with $N_{fr} = 350$ pulses (red and blue diamonds), and a rolling bunch pattern (black diamonds). The range of $\bar{\langle k\rangle} (q)$ for each attenuator setting is displayed as horizontal dashed lines. This range corresponds to the variation of $\bar{\langle k\rangle}$ along the different $q$-bins on the detector, and not fluctuations of the intensity. The error of $D_0$ is depicted as vertical, solid lines.

\begin{figure}[ht]
\begin{center}
\includegraphics[width=1\textwidth]{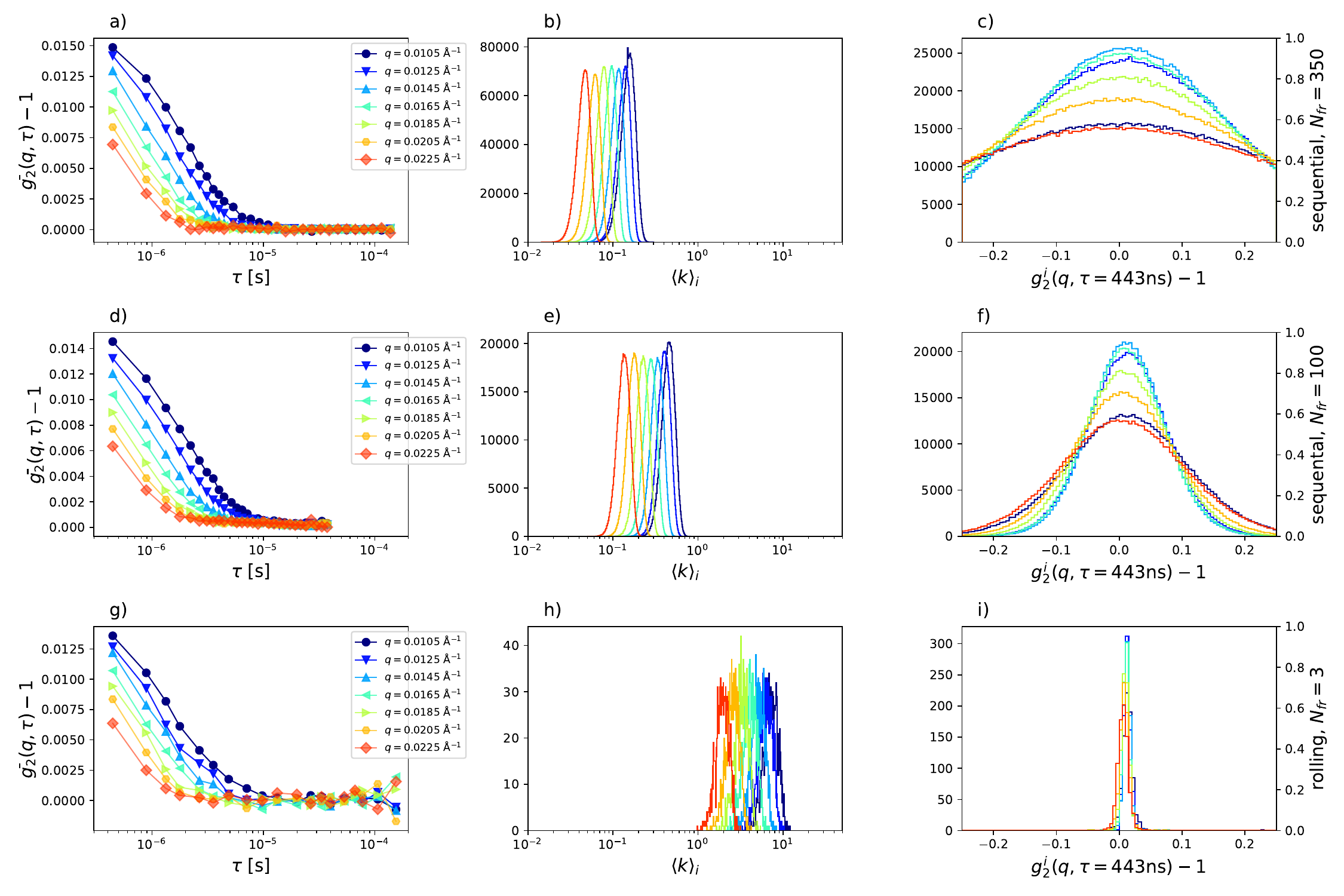}
\end{center}
\caption{Comparison of data obtained with sequential and rolling bunch pattern. The first line (a)-c)) shows data with the normal, sequential pulse pattern ($N_{fr} = 350$), the second line (d)-f)) with $N_{fr} = 100$ and the third line (g)-i)) the rolling bunch pattern with $N_{fr} = 3$. The first column (a), d), and g)) displays the measured correlation functions, the second column (b), e), and h)) the distribution of scattering intensities $\langle k \rangle_i$ and the third column (c), f), and i)) the distribution of $g_2^i(q, \tau)$. \label{fig:snr_comp}}
\end{figure}

To probe for deviations from equilibrium dynamics during the pulse train in the normal scheme, we evaluate the dynamics separately using only the first and the last 100 pulses of each train; these results are shown as blue and red data points, respectively. While $D_0$ remains approximately constant at low $\bar{\langle k\rangle}$, a noticeable increase is observed once $\bar{\langle k\rangle}$ exceeds a threshold value; the last data points within the "equilibrium regime" for each number of pulse are marked by arrows (corresponding to the three correlation functions shown in Fig. \ref{fig:snr_comp}). Notably, the onset occurs at lower $\bar{\langle k\rangle}$ for the later pulses (red) and at higher $\bar{\langle k\rangle}$ for earlier (blue), consistent with a stronger perturbation with a larger number of pulses. An even higher $\bar{\langle k\rangle}$ value can be reached using the rolling bunch pattern and the corresponding threshold appears about one order of magnitude higher, leaving the sparse data regime with $\bar{\langle k\rangle} > 1$.

The obtained threshold values are also added in Fig.~\ref{fig:snr_k_n} for direct comparison with the previous modeling. While the two data points stemming from the normal bunch pattern follow the sketched scaling very well, the black data point obtained with the rolling pattern suggests that the scaling is somewhat less steep than indicated by the red constant-dose-constraint lines. In fact, not only the dose but also the dose-rate was found to be a crucial property \cite{reiser2022, timmermann2023}, which is not considered in this simple picture so far. Nevertheless, we demonstrate experimentally that the accessible single-shot intensity $\langle k \rangle_i$ can in fact be increased by more than an order of magnitude, while still probing equilibrium dynamics.

The correlation functions obtained with the normal and rolling bunch pattern are also shown in Fig.~\ref{fig:snr_comp}, together with the corresponding histograms of single-shot intensities $\langle k \rangle_i$ and of single pulse-pair values $g_2^i(q,\tau)$. For the lower flux level, all 350 pulses are taken into account; consequently, the corresponding $\bar{g_2}(q, \tau)$ function spans a larger delay-time range as for $N_{fr} =100$. As shown before, the smaller $N_{fr}$, the higher the accessible $\langle k \rangle_i$ values. Correspondingly, the distribution of single pulse-pair $g_2^i(q,\tau)$ values becomes narrower, as expected from photon-statistics considerations. 

From these distributions, the single pulse-pair SNR can be calculated and
\begin{equation}
    \mathrm{SNR}_{pp}
    =
    \frac{
    \left\langle
    g_{2,\mathrm{on}}^i(q,\tau)-g_{2,\mathrm{off}}^i(q,\tau)
    \right\rangle_{fr}
    }{
    \sigma_{fr}\!\left[
    g_{2,\mathrm{on}}^i(q,\tau)-g_{2,\mathrm{off}}^i(q,\tau)
    \right]
    } .
    \label{eq:exp}
\end{equation}

\begin{sidewaystable}[p]
\centering
\caption{Comparison of measured and expected SNR for normal and rolling bunch pattern. \label{tab:snr_comp} }
\begin{tabular}{l l c c c c c c c c c c c c}
\toprule
$q$ & N$_f$ & $\langle k\rangle_i$ & $\langle k\rangle_i$ & CV & $N_{pix}$ & $g_2 - 1$ & $g_2 - 1$ & SNR (pp) & SNR (pp) & ratio & $(N_{fr}-1)N_{rep}$ & SNR & SNR \\

[\AA$^{-1}$] & (Mode) & mean & std & & & mean & std & calc. & exp. & exp./calc. & & overall & per min \\
\midrule
\multirow{3}{*}{0.010}
& 350 (seq) & 0.16 & 0.03 & 0.18 & 948 & 0.015 & 0.29 & 0.0492 & 0.0509 & 1.03 & 2201492 & 75.5 & 23.3 \\
& 100 (seq) & 0.46 & 0.08 & 0.17 & 948 & 0.015 & 0.10 & 0.1428 & 0.1465 & 1.03 & 623997  & 115.7 & 35.7 \\
& 3 (roll)  & 7.55 & 1.36 & 0.18 & 949 & 0.014 & 0.01 & 2.1574 & 1.2561 & 0.58 & 794     & 35.4 & 13.8 \\
\midrule
\multirow{3}{*}{0.012}
& 350 (seq) & 0.14 & 0.02 & 0.17 & 2799 & 0.014 & 0.19 & 0.0715 & 0.0751 & 1.05 & 2201492 & 111.4 & 34.3 \\
& 100 (seq) & 0.41 & 0.07 & 0.16 & 2799 & 0.013 & 0.07 & 0.1982 & 0.2007 & 1.01 & 623997  & 158.5 & 48.9 \\
& 3 (roll)  & 6.52 & 1.14 & 0.18 & 2797 & 0.013 & 0.01 & 2.9709 & 2.3087 & 0.78 & 794     & 65.1 & 25.3 \\
\midrule
\multirow{3}{*}{0.014}
& 350 (seq) & 0.11 & 0.02 & 0.16 & 4479 & 0.013 & 0.18 & 0.0685 & 0.0730 & 1.06 & 2201492 & 108.3 & 33.4 \\
& 100 (seq) & 0.34 & 0.06 & 0.16 & 4479 & 0.012 & 0.06 & 0.1895 & 0.1918 & 1.01 & 623997  & 151.5 & 46.7 \\
& 3 (roll)  & 5.29 & 0.92 & 0.17 & 4467 & 0.012 & 0.01 & 2.9584 & 2.2814 & 0.77 & 794     & 64.3 & 25.0 \\
\midrule
\multirow{3}{*}{0.016}
& 350 (seq) & 0.09 & 0.02 & 0.16 & 5949 & 0.011 & 0.18 & 0.0568 & 0.0614 & 1.08 & 2201492 & 91.2 & 28.1 \\
& 100 (seq) & 0.28 & 0.05 & 0.16 & 5951 & 0.010 & 0.06 & 0.1555 & 0.1601 & 1.03 & 623997  & 126.4 & 39.0 \\
& 3 (roll)  & 4.26 & 0.74 & 0.17 & 5954 & 0.011 & 0.01 & 2.5382 & 1.7353 & 0.71 & 794     & 48.9 & 19.0 \\
\midrule
\multirow{3}{*}{0.019}
& 350 (seq) & 0.08 & 0.01 & 0.16 & 7028 & 0.010 & 0.21 & 0.0432 & 0.0466 & 1.08 & 2201492 & 69.2 & 21.3 \\
& 100 (seq) & 0.23 & 0.04 & 0.16 & 7031 & 0.009 & 0.07 & 0.1187 & 0.1233 & 1.04 & 623997  & 97.4 & 30.0 \\
& 3 (roll)  & 3.39 & 0.58 & 0.17 & 7033 & 0.009 & 0.01 & 1.8423 & 1.3586 & 0.74 & 794     & 38.3 & 14.9 \\
\midrule
\multirow{3}{*}{0.020}
& 350 (seq) & 0.06 & 0.01 & 0.16 & 8544 & 0.008 & 0.24 & 0.0323 & 0.0347 & 1.07 & 2201492 & 51.5 & 15.9 \\
& 100 (seq) & 0.18 & 0.03 & 0.16 & 8544 & 0.008 & 0.08 & 0.0884 & 0.0916 & 1.04 & 623997  & 72.4 & 22.3 \\
& 3 (roll)  & 2.66 & 0.45 & 0.17 & 8532 & 0.008 & 0.01 & 1.4066 & 1.1893 & 0.85 & 794     & 33.5 & 13.0 \\
\midrule
\multirow{3}{*}{0.022}
& 350 (seq) & 0.05 & 0.01 & 0.16 & 9507 & 0.007 & 0.30 & 0.0216 & 0.0231 & 1.07 & 2201492 & 34.3 & 10.6 \\
& 100 (seq) & 0.14 & 0.02 & 0.16 & 9509 & 0.006 & 0.10 & 0.0587 & 0.0611 & 1.04 & 623997  & 48.3 & 14.9 \\
& 3 (roll)  & 2.01 & 0.34 & 0.17 & 9501 & 0.006 & 0.01 & 0.8561 & 0.7781 & 0.91 & 794     & 21.9 & 8.5 \\
\bottomrule
\end{tabular}
\end{sidewaystable}

The resulting values are provided in Table~\ref{tab:snr_comp} for $\tau = 443 \ \mathrm{ns}$: for the standard sequential acquisition one reaches again the expected SNR. Also, measurements with less pulses ($N_{fr} = 100$) reach consistently a higher SNR. This confirms the previous statement that indeed fewer pulses provide a higher SNR in the dose limited regime.

The highest overall $SNR_{pp}$ can be achieved using the rolling bunch pattern due to the higher attainable intensities $\langle k\rangle_i$. However, it can be also noted that the experimentally obtained $SNR_{pp}$ reaches only $60 - 90 \%$ of the expected values. This discrepancy may indicate the presence of additional experimental noise sources that become more relevant in the rolling bunch scheme implementation. Potential contributions from increased beam position fluctuations in the rolling bunch pattern can be excluded, since the speckle contrast $\beta$ obtained from the y-axis intercept of $\bar{g_2}(q, \tau)$ is identical for both the normal and rolling modes. Intensity fluctuations can also be ruled out as the origin of the discrepancy, as they are already incorporated in the theoretical expectation value and the coefficients of variation (CV) remain unchanged between both cases, as shown in Table~\ref{tab:snr_comp} as well. An additional noise contribution may arise from drifts in the AGIPD detector baseline between the trains used to calculate $g_{2,off}$. In sequential mode, these trains are separated by $100\,\mathrm{ms}$ (one train interval), compared with $1\,\mathrm{s}$ (10 train intervals) in rolling mode. The longer separation could allow larger baseline changes, reducing the effectiveness of the $g_{2,off}$ subtraction and introducing additional noise. This hypothesis could be tested by varying the number of patterns in the rolling cycle from the current 10, thereby changing the temporal separation, or by repeating the measurements with a lower-noise detector. The origin of the discrepancy nevertheless remains to be established.

Additionally, one can not only compare the SNR from a single pulse-pair but also the overall SNR that can be reached per minute of experiment. It is obvious from the distributions that much less pulse-pairs are sampled in case of the rolling bunch pattern, as $N_{fr}$ is smaller and $N_{rep}$ increases only with $1 \ \mathrm{s}^{-1}$. Still, due to the higher per frame intensities, comparable SNRs per minute of experiment can be achieved, which demonstrate the competitive performance of the rolling bunch pattern. These results show that the rolling bunch pattern can potentially benefit measurements of weakly scattering or highly radiation-sensitive samples, for which per-frame scattering intensities may be insufficient in standard sequential mode.

\section{Summary and outlook}
\label{sec:summary}

Burst-mode XFELs provide substantial flexibility for XPCS measurements with very high temporal resolution, while the pulse pattern can be adapted to optimize the use of the available photon budget. This flexibility creates opportunities for more dose-efficient XPCS measurements, especially for samples with low scattering cross sections or limited radiation tolerance. In this work, we demonstrated how this potential can be exploited using a rolling bunch pattern. This approach enabled substantially higher per-pulse intensities, while maintaining equilibrium dynamics in the investigated sample. Since the SNR of individual pulse-pair estimates can be substantially increased, the overall SNR per unit measurement time reaches close to that of the conventional sequential scheme in the present implementation. The additional experimental complexity must however be weighed against the achievable gain in per-pulse intensity for future applications and may still make the conventional sequential scheme preferable for sufficiently scattering and radiation-tolerant samples. Conversely, the rolling pattern is particularly beneficial for weakly scattering samples for which detector-imposed intensity limits would otherwise make the experiment impossible. The successful reconstruction of a full correlation function from acquisitions with varying delay times is additionally encouraging for future twin-bunch experiments, which require the construction of $g_2(q,\tau)$ from many double-pulse measurements.

Additionally, we tested an analytical framework for estimating the SNR in multispeckle X-ray photon correlation spectroscopy under dose-limited conditions. The agreement between the predicted and measured SNR for the conventional sequential acquisition supports the validity of this description. The analysis shows that, for a fixed dose, acquiring fewer frames at higher per-pulse intensity yields a higher SNR than distributing the dose over a larger number of low-intensity frames. Related strategies at synchrotron radiation sources have been recently demonstrated \cite{hoshino2026,chevremont2026} as well.

We further found that, for the Self-Amplified Spontaneous Emission (SASE) operation investigated at the European XFEL, pulse-to-pulse intensity fluctuations were sufficiently small to cause only a negligible reduction in the achievable SNR. The demonstrated dependence of the SNR on intensity fluctuations shows however that using monochromators with SASE radiation may be disadvantageous in certain cases, since the associated intensity fluctuations may outweigh the gain in speckle contrast. Consequently, inserting a monochromator can reduce the SNR even if it increases the contrast of the correlation function.

The developed framework is also relevant for evaluating advanced XFEL operation modes, such as Hard X-ray Self-Seeding \cite{liu2023cascaded}. In such cases, the bandwidth, which determines the longitudinal coherence length and contributes to the speckle contrast, is not the only important parameter. The pulse-to-pulse intensity stability must also be considered, since stronger intensity fluctuations can significantly reduce the SNR and potentially offset the benefit of increased contrast. This highlights the importance of characterizing both coherence and intensity stability when optimizing new XFEL modes. Looking further ahead, X-ray free-electron laser oscillators (XFELOs) could combine narrow bandwidth, and thus high longitudinal coherence, with stable pulse energies and beam pointing \cite{rauer2025}. Such a source could provide an almost ideal platform for coherent scattering experiments in the future.

\appendix 

\section{Modeling of intensity fluctuations}
\label{sec:model}

\begin{figure}[ht]
\begin{center}
\includegraphics[width=0.95\textwidth]{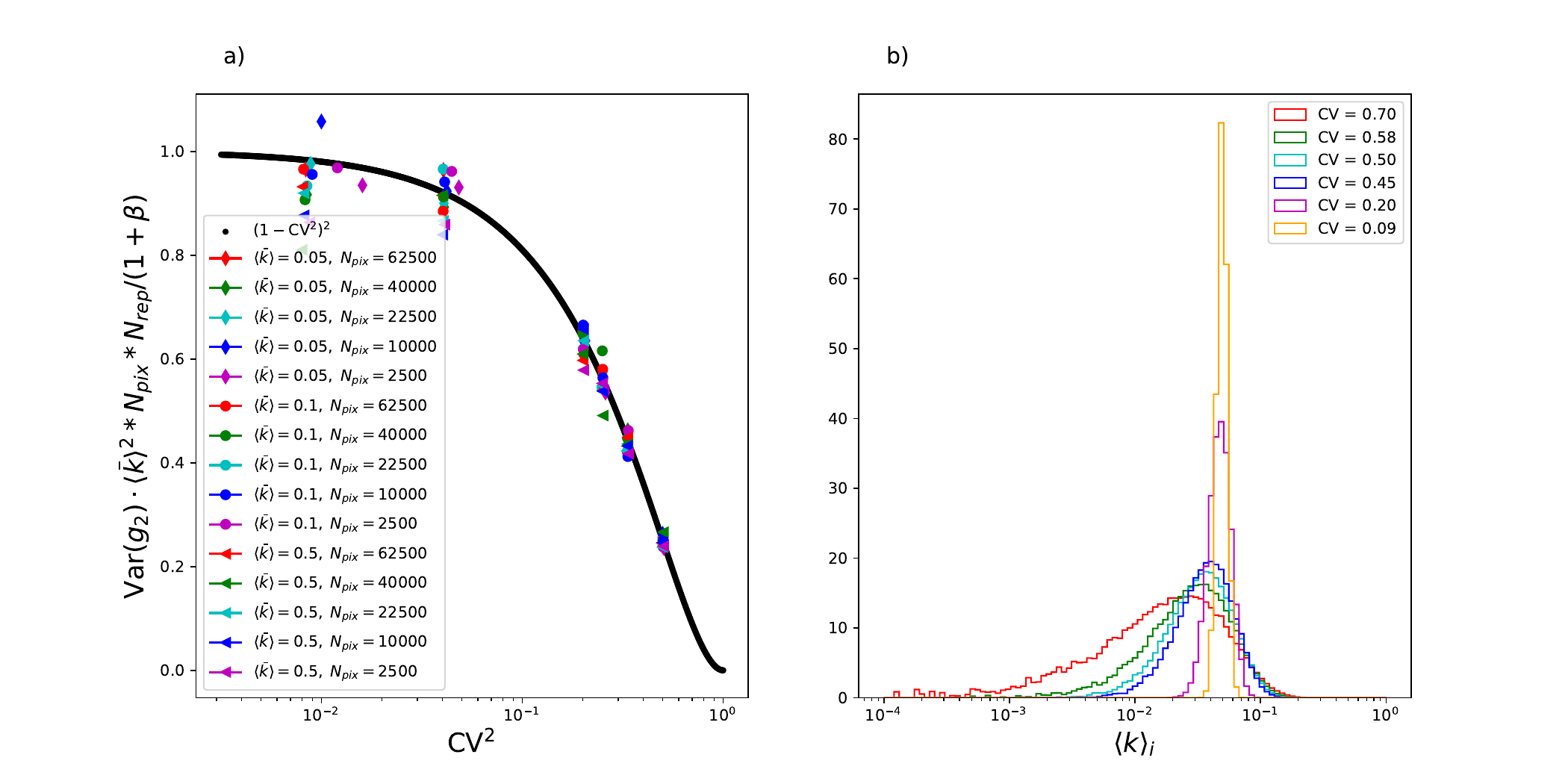}
\end{center}
\caption{a) Variance of $g_2$, normalized on the mean intensity $\langle k \rangle^2$, number of pixel $N_{pix}$, number of repetitions $N_{rep}$ and contrast $1+ \beta$, as a function of intensity fluctuations characterized by the coefficient of variation CV. The black line indicates the $(1-\mathrm{CV}^2)$ dependence. b) Single pulse intensity distributions in mean number of photons per pixel $\langle k \rangle_i$, shown for different values of CV. \label{fig:cv_numerical}}
\end{figure}

To verify Eq.~\ref{eq:snr_pulsed}, we perform numerical simulations of a series of speckle patterns, consisting of $N_{pix}$ detector pixels with a contrast of $\beta =0.05$. From the underlying speckle pattern, $N_{fr}$ detector frames were simulated by applying random Poisson sampling to represent photon counting statistics. In order to model temporal intensity fluctuations, the mean intensity $\langle k \rangle_i$ of each frame was drawn from a Gamma distribution with shape parameter $M$ and scale parameter $\bar{\langle k \rangle}/M$. This procedure generates controlled levels of pulse-to-pulse intensity fluctuations characterized by the coefficient of variation $\mathrm{CV}=1/\sqrt{M}$. The parameter $M$ was varied over the range $M \in [2,3,4,5,25,125]$, while the number of detector pixels was varied according to $N_{pix} \in [250 \times 250, 200 \times 200, 150 \times 150, 100 \times 100, 50 \times 50]$ and the mean intensity $\langle \bar{k} \rangle \in [0.05, 0.1, 0.5]$. For each parameter set, $N_{fr}=1000$ detector frames were generated and the simulation was repeated $N_{rep} = 100$ times to obtain sufficient statistics. The resulting variance of the $g_2^i(n = 1)$ is shown in Fig.~\ref{fig:cv_numerical} as a function of CV${}^2$ of the frame intensity distribution. By normalizing the variance to the different input parameters, one can see that the resulting term follows the $(1-\mathrm{CV}^2)$ dependence (black line) indicated by Eq.~\ref{eq:snr_pulsed}.

\section{Experimental details}
\label{sec:exp}

The measurements were performed at the Materials Imaging and Dynamics (MID) instrument \cite{madsen2021} of the European X-ray Free-Electron Laser Facility \cite{decking2020mhz} using the small-angle X-ray scattering (SAXS) configuration. The experiment employed SASE radiation from the SASE2 undulator directly at a photon energy of $9$ keV, without a monochromator and without self-seeding. The X-ray beam was focused onto the sample using the compound refractive lenses (CRL2). The focusing conditions were deliberately chosen to produce a beam size of approximately $20~\mu$m at the sample, rather than the smallest available focus. This comparatively large beam size reduced the incident fluence and thereby helped to minimize beam-induced heating and radiation damage, while retaining sufficient contrast for XPCS measurements.

The samples were filled in quartz capillaries and mounted on the fast solid sample scanner (FSSS) stage, which was operated inside the vacuum sample chamber. The FSSS translated the sample between successive pulse trains so that each train illuminated a fresh sample position, while the sample position remained fixed during the individual train. Scattered X-rays were recorded in the forward-scattering geometry with the AGIPD 1M detector \cite{allahgholi2019}. The individual scattering patterns from the pulses within a train were stored in separate AGIPD memory cells and read out between trains. The detector data were processed using the corresponding calibration data, including pedestal, gain, common-mode and bad-pixel corrections \cite{sztuk2024}.

Residual pedestal variations between AGIPD storage cells can generate spurious correlations. We therefore evaluated correlations between neighboring trains as off-correlations and subtracted them from the corresponding correlations between pulses within the same train. This procedure, described in detail in \cite{dallari2021analysis, leonau2025}, accounts for time-dependent variations of the detector pedestals and results in a corrected correlation function with a baseline close to zero rather than one. The same correction procedure was applied to the conventional sequential pulse pattern and to the rolling bunch-pattern measurements.

\section{A rolling bunch pattern scheme}
\label{sec:roll}
The rolling bunch pattern is implemented as a discrete set of $N_{pat} = 10$ predefined pulse configurations, each assigned to a consecutive pulse train and executed cyclically. Each configuration is defined by the indices of the electron bunches within a train that generate X-ray pulses ($p_i$), the corresponding separations in index $\Delta p$ and time $\Delta t$. Each pattern contains exactly three pulses and the temporal delays are determined by the bunch spacing. An overview of the actual values in shown in table \ref{tab:rolling_patterns}, which were chosen to approximate logarithmic spacing over the accessible time window.

\begin{table}[ht]
\centering
\caption{Cycled three-pulse patterns used for the rolling bunch scheme. Each train contains three pulses at indices ($p_1$,$p_2$,$p_3$) within the pulse train.}
\label{tab:rolling_patterns}
\begin{tabular}{c c c c}
\hline
Pattern ID & Pulse indices $
(p_1,p_2,p_3)
$ & $\Delta p$ & $\Delta t$ \\ \hline
1  & (80, 81, 82)   & $1$  & $443 \mathrm{ns}$ \\
2  & (265, 267, 269) & $2$  & $886 \mathrm{ns}$ \\
3  & (294, 297, 300) & $3$  & $1.33 \mathrm{\mu s}$ \\
4  & (132, 140, 148) & $8$  & $3.54 \mathrm{\mu s}$ \\
5  & (195, 206, 217) & $11$ & $4.87 \mathrm{\mu s}$  \\
6  & (114, 144, 174) & $30$ & $13.3 \mathrm{\mu s}$  \\
7  & (194, 239, 284) & $45$ & $19.9 \mathrm{\mu s}$  \\
8  & (102, 177, 252) & $75$ & $33.2 \mathrm{\mu s}$  \\
9  & (68, 188, 308)  & $120$ & $53.2 \mathrm{\mu s}$  \\
10 & (1, 176, 351)   & $175$ & $77.5 \mathrm{\mu s}$  \\
\hline
\end{tabular}
\end{table}

Each pulse is mapped to a specific AGIPD storage cell, so that each storage cell is illuminated only once per full cycle and that known problematic storage cells are excluded. Therefore, each illuminated storage cell is temporally bracketed by nine non-illuminations of the same cell before and nine after the train with signal. These measured dark values are used to estimate the local dark pedestal. Since the pedestal exhibits time-dependent fluctuations, its accurate reconstruction benefits from dark measurements taken as close in time as possible to the signal. The chosen configuration therefore represents the minimal achievable temporal separation between signal and dark references. The subsequent data analysis is detailed in sec. \ref{sec:rolling} of the main manuscript.

\begin{acknowledgements}
We acknowledge European XFEL in Schenefeld, Germany, for provision of X-ray free-electron laser beamtime at MID SASE2 under proposal number 3118 and would like to thank the staff for their assistance. This research was supported in part through the Maxwell computational resources operated at Deutsches Elektronen-Synchrotron DESY, Hamburg, Germany.
\end{acknowledgements}

\DataAvailability{Data recorded for the experiment at the European XFEL are available at doi:10.22003/XFEL.EU-DATA-003118-00.}

\bibliography{iucr.bib} 

\end{document}